\documentclass[journal=jacsat,manuscript=article]{achemso}
\usepackage[version=3]{mhchem} % Formula subscripts using \ce{}

\author{Yu.A. Budkov}
\affiliation{School of Applied Mathematics, Tikhonov Moscow Institute of Electronics and Mathematics, National Research University Higher School of Economics, Tallinskaya st. 34, 123458 Moscow, Russia}
\alsoaffiliation{G.A. Krestov Institute of Solution Chemistry of the Russian Academy of Sciences, Academicheskaya st., 1, 153045 Ivanovo, Russia}
\email{ybudkov@hse.ru}
\author{A.V. Sergeev}
\affiliation{N.N. Semenov Federal Research Center for Chemical Physics, Kosygina Street 4,
119991 Moscow, Russia}
\alsoaffiliation{Lomonosov Moscow State University, Leninskie gory 1, 119991 Moscow,
Russia}
\email{a.sergeev@chph.ras.ru}
\author{S.V. Zavarzin}
\affiliation{School of Applied Mathematics, Tikhonov Moscow Institute of Electronics and Mathematics, National Research University Higher School of Economics, Tallinskaya st. 34, 123458 Moscow, Russia}
\author{A.L. Kolesnikov}
\affiliation{Institut f\"{u}r Nichtklassische Chemie e.V., Permoserstr. 15, 04318 Leipzig, Germany}

\title[]{Two-component electrolyte solutions with dipolar cations on a charged electrode: Theory and computer simulations}

\keywords{Electric double layer, modified Poisson-Boltzmann equation, electrolytes, energy storage, energy conservation.}

\begin{document}

\begin{abstract}
The development of advanced electrochemical devices for energy conversion and storage requires fine tuning of electrode reactions, which can be accomplished by altering the electrode/solution interface structure. Particularly, in case of an alkali-salt electrolyte the electric double layer (EDL) composition can be managed by introducing organic cations (e.g. room temperature ionic liquid cations) that may possess polar fragments. To explore this approach, we develop a theoretical model predicting the efficient replacement of simple (alkali) cations with dipolar (organic) ones within the EDL. For the typical values of the molecular dipole moment ($2-4~D$) the effect manifests itself at the surface charge densities higher than 30 $\mu C/cm^2$. We show that the predicted behavior of the system is in qualitative agreement with the molecular dynamics simulation results. \end{abstract}

\section{Introduction}
The ongoing transition to ‘green’ energy urges us to develop advanced electrochemical devices for energy conversion and storage and, consequently, increases the need for a deeper theoretical understanding of such systems. The electrode/solution interface, which is an essential element of electrochemical devices, is especially interesting but presents a challenge for researchers. At the charged electrode/electrolyte interface the concentration of ions may be many times higher than the bulk solution value. That, of course, strongly affects the rate of the electrochemical reactions involving ionic species. Thus, tuning the electric double layer (EDL) composition may be of use in a number of electrochemical devices. For instance, in alkali-metal batteries (including Li-ion ones) the so-called solid-electrolyte interphase (SEI) layer is formed at the surface of both the anode and the cathode active materials. SEI consists of insoluble and partially soluble reduction products of electrolyte components and has a great impact on battery performance \cite{peled2017sei}. Particularly, the SEI structure affects the electron transfer kinetics and the battery cycle life. Controlling the EDL composition would be an extra tool for managing the SEI formation process. In an alkali-salt electrolyte the interface structure can be altered by introducing organic cations that would behave differently inside the EDL due to the factors arising from its complex molecular structure, e.g. steric limitations, dipole moment of the functional groups and specific interaction with the surface such as $\pi-\pi$ stacking. The remarkable modifiability of the organic cation implies that a compound can be designed to efficiently alter the EDL structure. Furthermore, the mixed electrolyte composition including both simple (alkali) and organic cations is already becoming more common as researchers and engineers are exploring the benefits of ionic liquids when applied to batteries, fuel cells, supercapacitors, solar cells, {\sl etc}.  \cite{fedorov2014ionic}. Thus, it is also becoming more relevant to make a theoretical description of the EDL incorporating both simple and organic cations.

Another bright example confirming the importance of the EDL structure is $Li$-air ($Li-O_2$) battery which is a promising electrochemical storage technology that hypothetically can provide several times higher specific energy than that of conventional $Li$-ion batteries \cite{aurbach2016advances}. Unfortunately, the cell capacity values achieved in the experiments so far are way below the theoretical expectations mostly due to the electrode surface passivation by the discharge product itself, i.e. $Li_2O_2$. There are two possible pathways of $Li_2O_2$ formation during the discharge process \cite{johnson2014role}. Relatively large micron-size $Li_2O_2$ particles which do not consume much of the electrode surface can be formed as a result of the chemical disproportionation reaction of $LiO_2$ intermediate in the bulk electrolyte \cite{zhai2013disproportionation, mitchell2013mechanisms}. Alternatively, it is possible to grow a thin $Li_2 O_2$ film through $Li O_2$ electrochemical reduction \cite{wen2013situ, johnson2014role} at the surface with the consumption of extra $Li^{+}$ ions. The predominance of the second surface-mediated pathway leads to complete electrode surface passivation (and cell “death”) while most of the available space inside the porous cathode is still unfilled with the discharge product. That drastically decreases the cell specific capacity. Exclusion of $Li^{+}$ ions from the EDL could inhibit the undesired surface-mediated pathway and slow down the passivation process thus enabling the accumulation of a larger amount of the discharge product inside the porous electrode. Introduction of organic cations into the electrolyte solution would be an easy practical way of controlling the EDL structure.

From the physics point of view, such organic cations should be attracted by the cathode more strongly than $Li^{+}$ ions. 
Thus, the greater attractive force acting on each of the organic cations must cause the lithium cations to be expelled from the cathode surface at its sufficiently large surface charge density. One of the possible ways to make the attraction of the organic cations stronger is using of an organic salt with highly polar cations. It can be, for instance, some room temperature ionic liquid, whose organic cations usually possess a rather high dipole moment or electronic polarizability. As is well known, in an inhomogeneous electric field a polar particle suffers a dielectrophoretic force (see, for instance, \cite{jones1979dielectrophoretic}) directed to the highest electric field. In our case, the electric field inhomogeneity takes place due to the electrostatic screening of the electrode charge by the mobile electrolyte ions, so that the maximal electric field is reached on the electrode surface. Therefore, the occurrence of a dipole moment on the organic cation should lead to its higher attraction to the electrode with respect to $Li^{+}$. The latter will provoke the expelling of lithium cations from the electrode surface  due to the excluded volume interactions and their complete replacement with the electrochemically inactive organic cations at a sufficiently large electrode surface charge density (see Fig. \ref{Schematic}). 

\begin{figure}[h!]
\center{\includegraphics[width=0.8\linewidth]{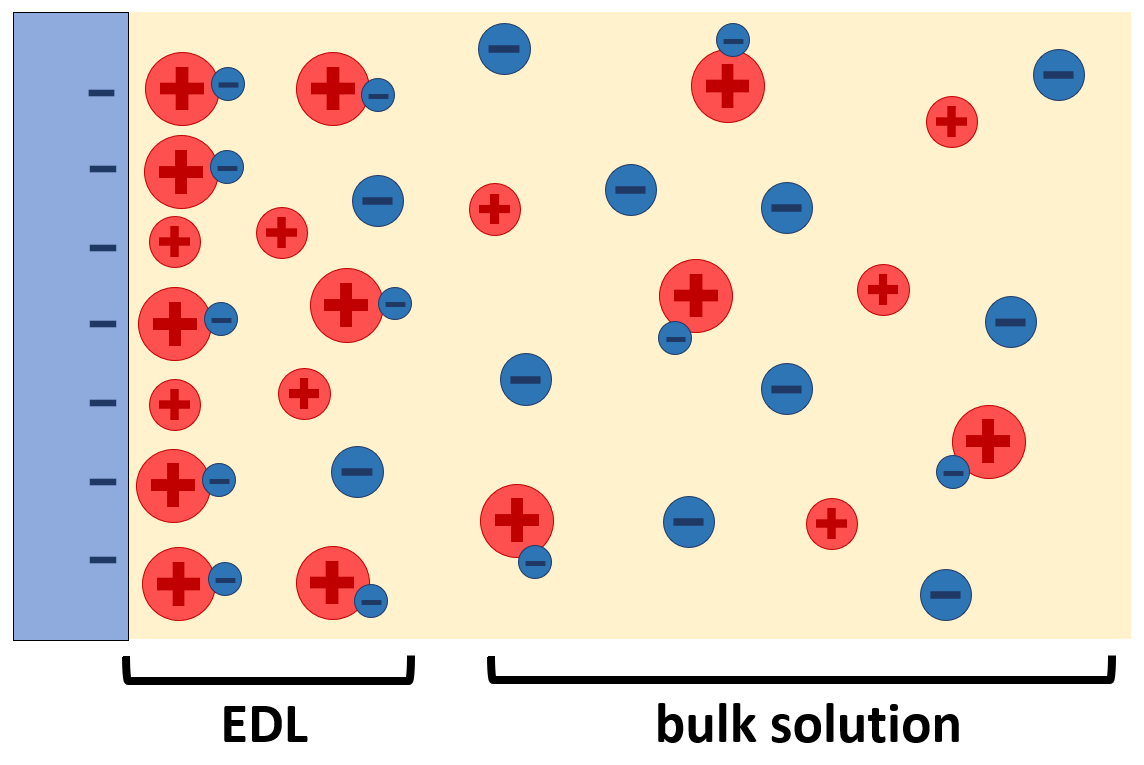}}
\caption{Schematic representation of the electrode/solution interface in mixed electrolyte containing simple (red circle) and molecular (large red circle with a small blue one) cations: bulk solution – dipole moments of molecular cations are disordered, equimolar cations concentration; EDL – dipoles are highly ordered according to electric field, molecular cations dominate.}
\label{Schematic}
\end{figure}

In order to understand whether it is possible to replace (at least partially) the lithium cations with organic dipolar cations on the cathode surface at physically reasonable parameters of the system, such as surface charge density, dielectric permittivity of the solvent, dipole moment of the cation, and ionic concentrations in the bulk, it would be useful to have a simple analytical model of a two-component electrolyte solution with one type of anions and two types of cations on the negatively charged metal electrode. Within such a theory, one of the cations would be a simple structureless particle with a charge $+e$, whereas the other one -- a particle possessing not only a charge $+e$, but also a certain permanent dipole moment $p$. A self-consistent field (SCF) theory, based on the modified Poisson-Boltzmann equation for the electrostatic potential and taking into account the steric interactions of ions and polarity of one type of cations could be such a theory. Despite the fact that there are a lot of SCF theories of electrolyte solutions taking into account polarizable/polar additives \cite{coalson1996statistical,abrashkin2007dipolar,gongadze2011langevin,frydel2011polarizable,budkov2018nonlocal,budkov2019statistical,Sin2017,Misra2013,Hatlo2012,Ben-Yaakov2011,Wei1993,Das2012,Jiang2014,gongadze2011generalized,budkov2016theory,budkov2018theory,budkov2015modified,nakayama2015differential,lopez2018diffuse,lopez2018multiionic}, to the best of our knowledge, up to present, the issues raised above have not been systematically discussed in the literature. Therefore, below we will formulate such a SCF theory and study in its framework the possibility, in principle, to exclude simple cations (which mimic the $Li^{+}$ ions) from the cathode surface by adding some dipolar cations (which mimic organic cations) to the electrolyte solution. In order to confirm our theoretical observations, we will compare the obtained results with those obtained by molecular dynamics simulation.

\section{Theory}
Let us consider an interface between a flat charged electrode with a surface charge density $\sigma$ and an electrolyte solution with one type of monovalent anions and two types of monovalent cations. Let us also assume that the cations of the first type (simple cations) carry an electric charge $e$, whereas the cations of the second type (molecular cations) in addition to the same charge possess a permanent dipole moment $p$. We start from the grand thermodynamic potential of the electrolyte solution in the local density approximation, which can be written in the form \cite{budkov2016theory,budkov2018theory}
\begin{equation}
\label{Grand_pot}
\Omega=\int\limits_{0}^{\infty}\left(-\frac{\varepsilon\varepsilon_0\mathcal{E}^2}{2}+ \rho_{c}\psi-c_{+}^{(2)}\Psi +f-\mu_{+}^{(1)}c_{+}^{(1)}-\mu_{+}^{(2)}c_{+}^{(2)}-\mu_{-}c_{-}\right)dz,
\end{equation} 
where $f=f(T,c_{+}^{(1)},c_{+}^{(2)},c_{-})$ is the Helmholtz free energy density of a reference system (a system without electrostatic interactions) and the auxiliary function
\begin{equation}
\Psi(z)=k_{B}T\ln\frac{\sinh{\beta p\mathcal{E}(z)}}{\beta p\mathcal{E}(z)},
\end{equation}
is introduced. $\mathcal{E}(z)=-\psi^{\prime}(z)$ is the local electric field and $\rho_{c}(z)=e\left(c_{+}^{(1)}(z)+c_{+}^{(2)}(z)-c_{-}(z)\right)$ is the local charge density of the ions; $T$ is the temperature and $k_B$ is the Boltzmann constant. Using the local Legendre transformation \cite{maggs2016general,budkov2016theory,budkov2018theory,mceldrew2018theory}, we can rewrite the grand thermodynamic potential in the form
\begin{equation}
\label{Grand_pot_2}
\Omega=-\int\limits_{0}^{\infty}dz\left(\frac{\varepsilon\varepsilon_0\mathcal{E}^2(z)}{2}+P(T,\bar{\mu}_{+}^{(1)},\bar{\mu}_{+}^{(2)},\bar{\mu}_{-})\right),
\end{equation}
where the local chemical potentials of the species
\begin{equation}
\bar{\mu}_{+}^{(1)}=\mu_{+}^{(1)}-e\psi,~\bar{\mu}_{+}^{(2)}=\mu_{+}^{(2)}-e\psi+\Psi,
\end{equation}
\begin{equation}
\bar{\mu}_{-}=\mu_{-}+e\psi
\end{equation}
are introduced. A variation of functional (\ref{Grand_pot_2}) with respect to the potential $\psi(z)$ leads to the following self-consistent field equation \cite{budkov2016theory}:
\begin{equation}
\frac{d}{dz}\left(\epsilon(z)\psi^{\prime}(z)\right)=-e\left(\bar{c}_{+}^{(1)}(z)+\bar{c}_{+}^{(2)}(z)-\bar{c}_{-}(z)\right), \end{equation}
where
\begin{equation}
\bar{c}_{+}^{(1,2)}=\frac{\partial{P}}{\partial{\bar{\mu}}_{+}^{(1,2)}}~,\bar{c}_{-}=\frac{\partial{P}}{\partial{\bar{\mu}}_{-}},
\end{equation}
are the local equilibrium concentrations of the ions and
\begin{equation}
\epsilon(z)=\varepsilon\varepsilon_0+\frac{p^2}{k_{B}T}\frac{L(\beta p\mathcal{E})}{\beta p\mathcal{E}}\bar{c}_{+}^{(2)}(z)
\end{equation}
is the local dielectric permittivity; $L(x)=\coth{x}-x^{-1}$ is the Langevin function. The first integral of the self-consistent field equation determining the condition of the solution mechanical equilibrium \cite{budkov2016theory,budkov2015modified} takes the form
\begin{equation}
\label{mech_eq_cond}
-\frac{\varepsilon\varepsilon_0\mathcal{E}^2}{2}-p\mathcal{E}\bar{c}_{+}^{(2)}L(\beta p\mathcal{E})+P(T,\bar{\mu}_{+}^{(1)},\bar{\mu}_{+}^{(2)},\bar{\mu}_{-})=P(T,{\mu}_{+}^{(1)},{\mu}_{+}^{(2)},{\mu}_{-}).
\end{equation}
Solving equation (\ref{mech_eq_cond}) with respect to $\mathcal{E}$ at different $\psi$, we obtain the implicit function $\mathcal{E}=\mathcal{E}(\psi)$. In order to obtain the potential profile $\psi(z)$, it is necessary to solve the ordinary first-order differential equation $\psi^{\prime}(z)=-\mathcal{E}(\psi)$ with the initial condition $\psi(0)=\psi_0(\sigma)$, where $\psi_0(\sigma)$ is the surface potential of the electrode corresponding to the fixed surface charge density $\sigma$. In order to obtain the function $\psi_0(\sigma)$, it is necessary to use the boundary condition $\epsilon_s\mathcal{E}_0=\sigma$,
where the local dielectric permittivity at the electrode is $\epsilon_s=\epsilon(0)$ and the electric field at the electrode $\mathcal{E}_{0}=\mathcal{E}(0)$ are introduced. The latter is an implicit function of the surface potential $\psi_0$, determined by eq. (\ref{mech_eq_cond}).

As a reference system, let us consider three-component symmetric lattice gas model \cite{maggs2016general,budkov2016theory,kornyshev2007double} with the equation of state
\begin{equation}
P(T,{\mu}_{+}^{(1)},{\mu}_{+}^{(2)},{\mu}_{-})=\frac{k_B T}{v}\ln\left(1+e^{\beta\mu_{+}^{(1)}}+e^{\beta\mu_{+}^{(2)}}+e^{\beta\mu_{-}}\right), 
\end{equation}
where $\beta=(k_{B}T)^{-1}$. The local concentrations of the ions within the three-component lattice gas model are determined by the following relations
\begin{equation}
\bar{c}_{+}^{(1)}(z)v=\frac{e^{\beta(\mu_{+}^{(1)}-e\psi(z))}}{1+e^{\beta(\mu_{+}^{(1)}-e\psi(z))}+e^{\beta(\mu_{+}^{(2)}-e\psi(z)+\Psi(z))}+e^{\beta(\mu_{-}+e\psi(z))}},
\end{equation}
\begin{equation}
\bar{c}_{+}^{(2)}(z)v=\frac{e^{\beta(\mu_{+}^{(2)}-e\psi(z)+\Psi(z))}}{1+e^{\beta(\mu_{+}^{(1)}-e\psi(z))}+e^{\beta(\mu_{+}^{(2)}-e\psi(z)+\Psi(z))}+e^{\beta(\mu_{-}+e\psi(z))}},
\end{equation}
\begin{equation}
\bar{c}_{-}(z)v=\frac{e^{\beta(\mu_{-}+e\psi(z))}}{1+e^{\beta(\mu_{+}^{(1)}-e\psi(z))}+e^{\beta(\mu_{+}^{(2)}-e\psi(z)+\Psi(z))}+e^{\beta(\mu_{-}+e\psi(z))}}.
\end{equation}

Using the electrical neutrality condition for the bulk solution $c_{-,b}=c_{+,b}^{(1)}+c_{+,b}^{(2)}$ and 
introducing the bulk concentrations of electrolytes $c_{1}=c_{+,b}^{(1)}$ and $c_{2}=c_{+,b}^{(2)}$, we obtain the relations for the species chemical potentials
\begin{equation}
\mu_{+}^{(1,2)}=k_{B}T\ln\left(\frac{c_{1,2}v}{1-2v(c_{1}+c_{2})}\right),~\mu_{-}=k_{B}T\ln\left(\frac{(c_{1}+c_{2})v}{1-2v(c_{1}+c_{2})}\right).
\end{equation}
\section{Computational details: Molecular dynamics simulations}
All the molecular dynamics simulations were performed with the help of the LAMMPS \cite{plimpton1993fast} simulation package. The cutoff distance for the electrostatic and van der Waals interactions was set to $2~nm$. The long range electrostatic interactions were calculated using the pppm scheme \cite{hockney1988computer}. The simulations we carried out in an $NVT$-ensemble at $300~K$ using a Langevin thermostat (dump parameter $100~fs$) wan integration time-step of $1~fs$. Periodic boundary conditions were applied along the $x$ and $y$ directions, while the unwanted interactions between the replicas along the $z$ direction were eliminated by using special algorithms provided within the LAMMPS package 
\cite{yeh1999ewald,ballenegger2009simulations}.
The simulation cell contained two parallel plates (parallel to $xy$ plane) constructed from frozen atoms ordered according to the hexagonal $2-d$ lattice with the parameter $a = 0.293~nm$ (that resembles Au(111) surface topology). The plate size was $58.6 \times 60.90$ $\AA$ ($xy$) with an $8~nm$ distance (along the $z$ axis) between them. The plates were charged oppositely by setting the required atomic charges. The space between the ‘electrode plates’ was filled with ions. The solvent effect was taken into account implicitly by setting the relative dielectric permittivity to 40 (close to that of commonly used organic solvents such as acetonitrile, dimethylsulfoxide, {\sl etc.}). The solvent viscosity effect was simulated by a Langevin thermostat. The van der Waals interactions were described by the Lennard-Jones 6-12 potential with the depth $\varepsilon=0.1~Kcal/mol$ and the repulsion distance equal to $0.3~nm$ and $0.38~nm$ for the ions and ‘electrode’ atoms, respectively. The simple cations (and anions) were represented by individual atoms with the charge $+1e$ ($-1e$) and mass $10~g/mol$. The organic cations were designed in a Drude-particle fashion with a constant dipole moment. It was composed of a Drude-core (mass $9.6~g/mol$, charge $+3e$) and a pseudo electron (mass $0.4~g/mol$, charge $-2e$). The distance between the core and the ‘electron’ was kept fixed during the simulation at the value of $0.4~\AA$ or $0.2~\AA$ that yield the dipole moment about $4~D$ ($3.84$) and $2~D$ ($1.92$), respectively. The pseudo-electrons were thermalized separately \cite{jiang2011high} at $T=1~K$ temperature by a special Langevin thermostat with a dump parameter of $20~fs$. 

At higher surface charge values the EDL capacity turned out to exceed the number of ions between the plates at the desired concentration ($1.0$ or $0.5~M$ of the cation-anion pairs). Therefore, prior to the main simulation an iterative procedure was employed to continuously add (or remove, in case of overshoot) ion pairs with cations of the required type to achieve the target bulk concentration value. \footnote{Here ‘bulk’ concentration is attributed to the central $60~\AA$ thick slab.} After the EDL was completely saturated and the target bulk concentrations were reached, the refiling procedure was stopped. The following productive run was performed to obtain the $1~ns$ long trajectory.       

\begin{figure}[h!]
\center{\includegraphics[width=0.8\linewidth]{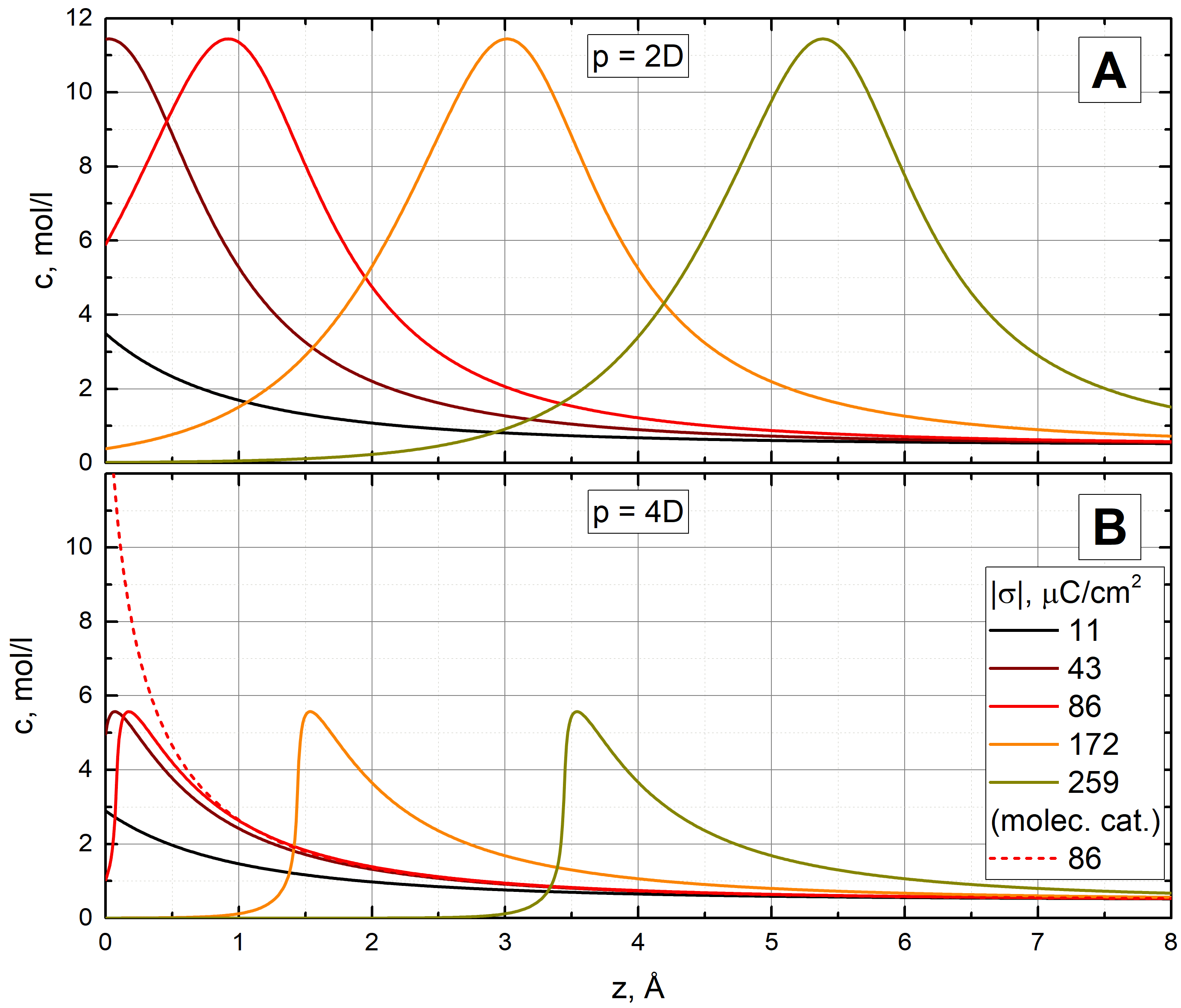}}
\caption{The concentration profiles of the simple cations calculated within the mean field theory at different values of the cathode surface charge density. The concentration profile of the molecular cations is also shown only for one case. The data are shown for the effective size $v^{1/3}=3.7~\AA$, temperature $T=300~K$, bulk concentrations of the cations $c_{1}=c_{2}=0.5~M$, and two dipole moments of the molecular cations $p=2~D$ (A) and $p=4~D$ (B).}
\label{Profiles_1}
\end{figure}

\section{Results and discussions}
First, we discuss the cation concentration profiles near the electrode surface, which can be predicted by the formulated above mean field theory at different values of dipole moment of the molecular cation $p$ and the cathode surface charge density $\sigma <0$. Let us consider the equimolar electrolyte compostion, i.e. $c_{1}=c_{2}=0.5~M$ (total $1~M$ salt concentration). Such value is within the typical range ($0.1-1.0~M$) for the real electrochemical devices discussed in the introductory section. We also assume that the dielectric permittivity $\varepsilon=40$ that approximately corresponds to the value of common organic solvents such as dimethylsulfoxide or acetonitrile. Figure \ref{Profiles_1} (A,B) shows the theoretical concentration profiles of the simple cations calculated for two values of the dipole moments of molecular cations at different values of the cathode surface charge density $|\sigma|$. As is seen, at a sufficiently small $|\sigma|$, the concentration of the simple cations monotonically decreases together with the distance from the electrode, so that its maximal value is attained at the electrode surface (at $z=0$). Moreover, higher absolute value of the surface charge density $|\sigma|$ causes the increase in the local concentration of the simple cations on the electrode. However, when the $|\sigma|$ exceeds a certain threshold value, the local concentration monotonically decreases, while the concentration profiles exhibit a pronounced maximum at a certain  distance $z>0$. This maximum can be shifted to higher distances by a further increase in the surface charge density. As is seen in Fig. \ref{Profiles_1} (B), the described effects become stronger at a larger dipole moment of the molecular cations. Moreover, at a rather large dipole moment of the molecular cations the maxima on the concentration profiles become sharper. Fig. \ref{Profiles_1} (B) also shows the concentration profile of the molecular cations only for one case. As is seen, at $|\sigma|=86~\mu C/cm^2$ the simple cations are depleted near the electrode surface, while the concentration of the molecular cations reaches the maximal value there. The latter clearly demonstrates the replacement of the simple cations by the molecular cations.    

\begin{figure}[h!]
\center{\includegraphics[width=0.8\linewidth]{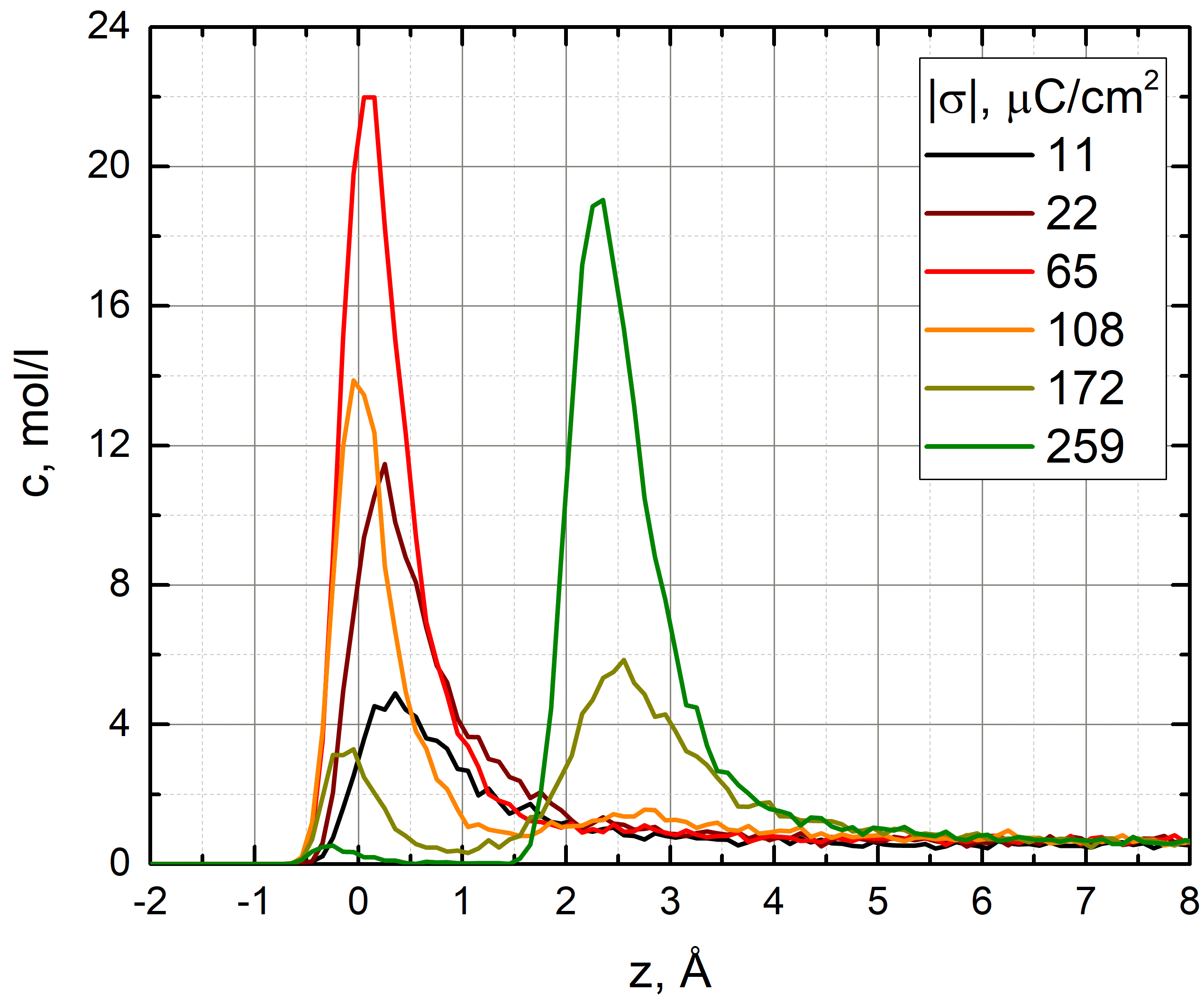}}
\caption{The concentration profiles of the simple cations obtained by the MD simulations for the different cathode surface charge density. The data are shown for $p=4~D$.}
\label{Profiles_2}
\end{figure}

\begin{figure}[h!]
\center{\includegraphics[width=0.8\linewidth]{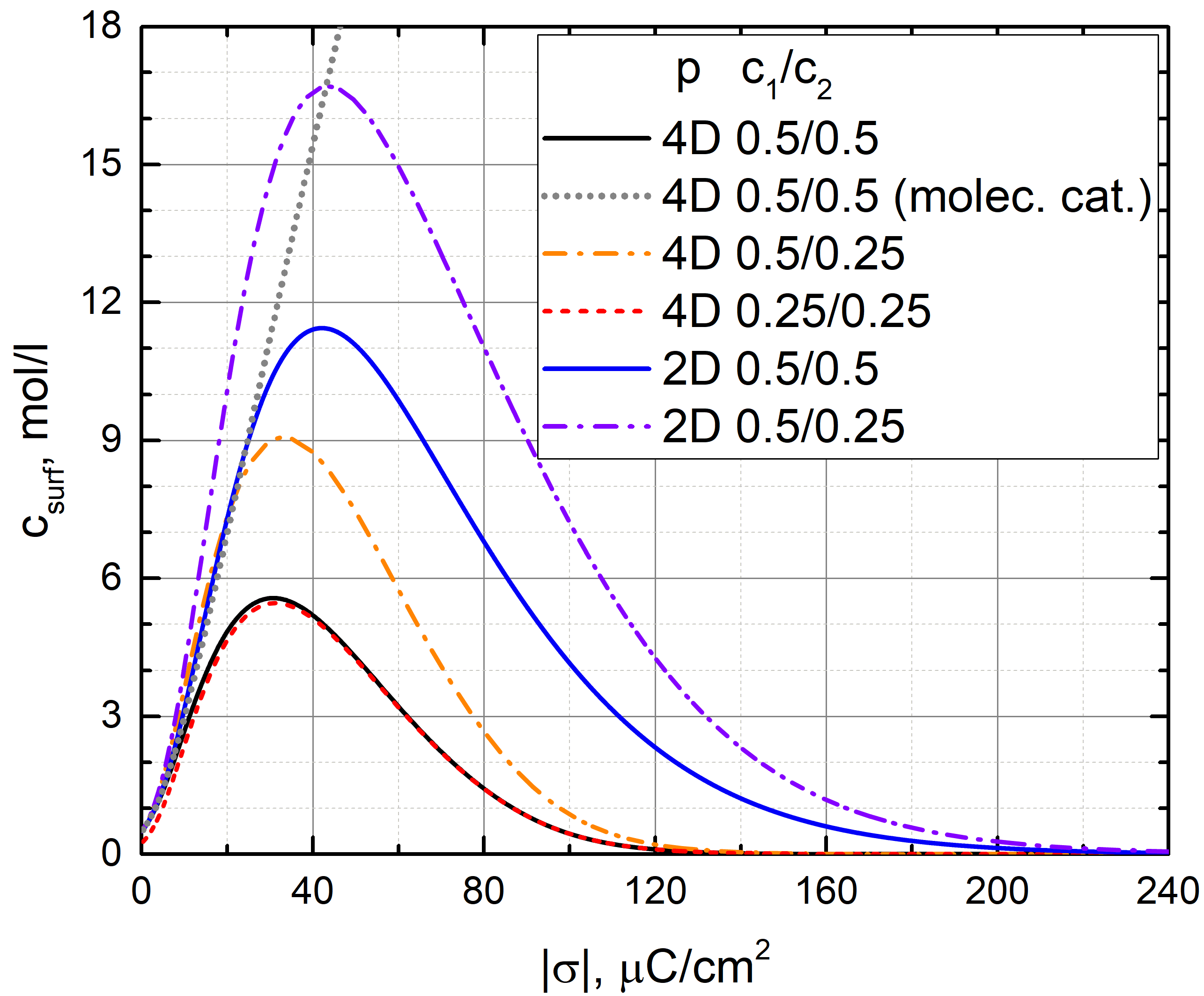}}
\caption{The local concentration of the simple cations on the electrode as a function of the cathode surface charge density calculated within the mean field approximation. The local concentration of molecular cations is also shown for only one case. The data are shown for different dipole moments and concentrations of the simple and molecular cations in the bulk.}
\label{surf_conc_1}
\end{figure}

\begin{figure}[h!]
\center{\includegraphics[width=0.8\linewidth]{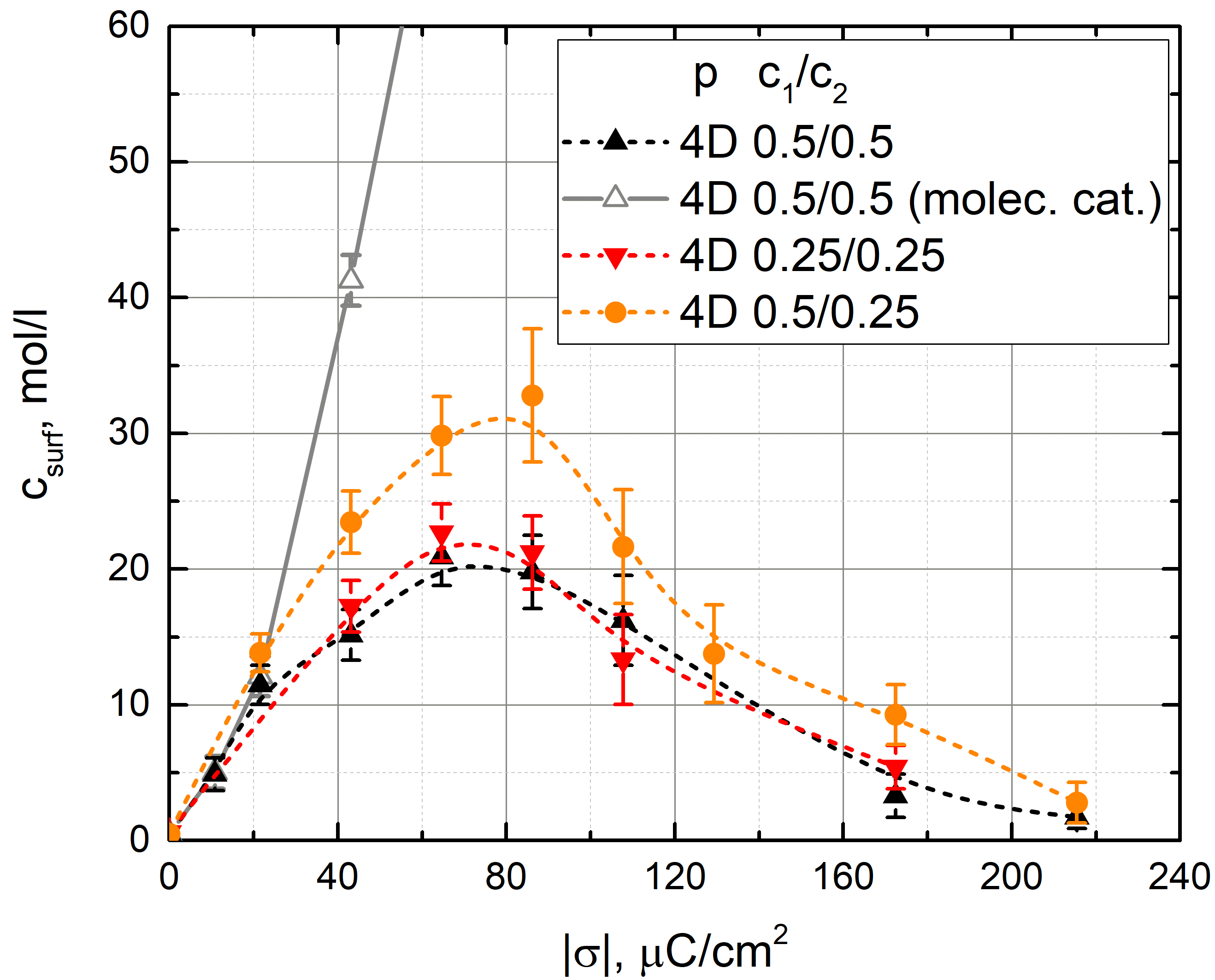}}
\caption{The local concentration of the simple cations on the electrode as a function of the cathode surface charge density obtained by the MD simulations. The local concentration of molecular cations is also shown for only one case. The data are shown for dipole moment $p=4~D$ of the molecular cation and different bulk concentrations of all cations.}
\label{surf_conc_2}
\end{figure}

Such a behavior can be interpreted as follows. At a rather small $|\sigma|$ there is a lot of free space for both types of the cations near the electrode surface, a higher surface charge density increases the local concentration of the simple cations on the electrode, while the concentration profiles are described by the monotonically decreasing functions of the distance. In this case, the maximum of concentration is reached on the electrode surface. At a sufficiently strong surface charge density the molecular cations are more strongly attracted to the electrode than the simple cations due to the occurrence of an additional dielectrophoretic force \cite{jones1979dielectrophoretic,budkov2018theory,budkov2016theory} acting on each dipolar cation. Therefore, at a rather large $|\sigma|$ the presence of the molecular cations near the electrode is more thermodynamically favorable, so that the simple cations are expelled from the $"$near-surface$"$ layer due to the steric interactions. The latter explains the occurrence of a pronounced maximum on the concentration profiles of the simple cations, which can be shifted by the increase in the surface charge density (see Figs. \ref{Profiles_1} (A,B)). The shift in the concentration maximum due to the increase in the surface charge density can also be explained by the simple cations exclusion the layers nearest to the electrode and their replacement with molecular cations. We would like to note that similar replacement from the cathode surface of alkaline cations by another alkaline cations due to a difference in their excess polarizabilities in aqueous medium was discussed within the mean-field theory in paper \cite{lopez2018diffuse,lopez2018multiionic}.

In order to confirm our theoretical findings, we performed MD simulations (for the details of the simulation setup, see the previous section). Fig. \ref{Profiles_2} shows the concentration profiles of the simple cations obtained by means of the MD simulations for the case $p=4~D$. The first concentration peak (which can be considered as a local concentration at the electrode surface) grows as the surface charge density increases up to about 65 $\mu C/cm^2$. A further increase in the electrode surface charge density leads to a decrease in the first concentration peak and simultaneous growth in the second one. Thus, the simple cations are expelled into the second ionic layer. The difference in comparison with the theoretical model is that the simple cations are pushed away from the surface not continuously (along the z coordinate) but layer by layer. Nevertheless, the general behavior of the concentration profiles predicted by the theoretical model is in qualitative agreement with that obtained by MD simulations. 

Let us now analyze the local concentration of the simple cations at the electrode surface as a function of the surface charge density for several electrolyte compositions. The results obtained within our mean field theory are presented in Fig. \ref{surf_conc_1}. The grey dotted line represents the surface concentration of the molecular cations. Comparing it with the simple cations concentration in the same case (black line) one can see how the surface ionic composition depends on the electrode charge density. Considering the equimolar compositions ($c_{1}/c_{2}=0.5/0.5$, i.e. $1~M$ total and $c_{1}/c_{2}=0.25/0.25$, i.e. $0.5~M$ total), it should be noted that the surface concentration of the simple (as well as molecular) cations weakly depends on the total bulk electrolyte concentration. The higher dipole moment of the molecular cations leads to a lower surface concentration of the simple cations, as the dielectrophoretic force is stronger in that case and the molecular cations are more strongly attracted to the cathode surface. The decline from the equimolar compositions in case of the higher bulk concentration of the simple cations ($c_{1}/c_{2}=0.5/0.25$, i.e. $0.75~M$ total) leads to higher surface concentration of simple cations, as there are less molecular cations to compete for the place within the EDL. However, the first ionic layer still consists mostly of molecular cations. It should be also noted that the position of the maximum of the function $c_{surf}=c_{surf}(\sigma)$ depends on the dipole moment $p$, but does not depend on the bulk electrolyte composition. Fig. \ref{surf_conc_2} presents the surface concentration dependences for $p = 4~D$ obtained by means of MD simulations. Although there are some quantitative differences, the qualitative behavior of the system is in good agreement with the theoretical predictions.

% The behavior is the same as of a system containing a common binary electrolyte, thus it is not related to the introduction of molecular cation. 

We would like to note that in the present theoretical mean field model as well as in the MD simulations we modelled the solvent as a continuous dielectric medium with constant dielectric permittivity. Thereby, we neglected the effects of solvent polarization causing a reduction in the dielectric permittivity near the charged electrode relative to the bulk solution \cite{yeh1999dielectric,gongadze2011langevin,gongadze2011generalized}. It is clear that accounting for this effect will not change qualitatively the simple cations exclusion effects, but can still significantly change the region of the surface charge densities, where this effect may take place. We neglected also an influence of ions on the dielectric permittivity of solution (so-called, dielectric decrement) \cite{Ben-Yaakov2011}. Moreover, in this study we did not take into account the short-range specific interactions between the ionic species and the electrode \cite{budkov2018theory,goodwin2017mean}. However, in this study we aimed to investigate the possibility, in principle, of excluding simple cations from the charged cathode by adding an organic salt with dipolar cations to the electrolyte solution at the simplest level of the theoretical model. The reasonable values of the surface charge densities, bulk concentrations of the electrolytes, and dipole moment of the organic cations, predicted by our mean field theory and qualitative agreement of the theoretical predictions with the MD simulations support the correctness of the obtained results.

\section{Conclusions}
In this work, we have formulated a mean field theory of the flat electric double layer on the interface between a metallic cathode and an electrolyte solution with two types of monovalent cations. One type of the cations is described as structureless charged particles (simple cation, e.g. alkali ion), whereas another one – as charged particle carrying polar groups with a considerable dipole moment (molecular cations). In practice the latter can be an organic cation (e.g. quaternary ammonium $NR_4^{+}$) with an attached acceptor group, such as trifluoromethyl ($-CF_3$), nitrile ($-CN$), carbonyl ($-COOR$), sulfonyl ($-SO_3 R$), {\sl etc}. We have shown that at a sufficiently high absolute value of the cathode surface charge density (about $30~\mu C/cm^2$), the simple cations are expelled from the cathode surface by the molecular cations due to the stronger attraction of the latter to the electrode. We have shown that the stronger attractive force is realized due to the occurrence of the dielectrophoretic force acting on each molecular cation in the inhomogeneous electrostatic field. We have confirmed our theoretical predictions by molecular dynamics simulations. The reported study estimates the strength of the cation replacement effect for various sets of parameters, which can be further used for experiment design.

\begin{acknowledgement}
The reported study was partially supported by the RFBR according to research project no. 18-31-20015. The MD simulation part of the study carried out by Artem Sergeev was supported by Russian Science Foundation grant no. 19-43-04112.
\end{acknowledgement}

%%%%%%%%%%%%%%%%%%%%%%%%%%%%%%%%%%%%%%%%%%%%%%%%%%%%%%%%%%%%%%%%%%%%%
%% The same is true for Supporting Information, which should use the
%% suppinfo environment.
%%%%%%%%%%%%%%%%%%%%%%%%%%%%%%%%%%%%%%%%%%%%%%%%%%%%%%%%%%%%%%%%%%%%%

%%%%%%%%%%%%%%%%%%%%%%%%%%%%%%%%%%%%%%%%%%%%%%%%%%%%%%%%%%%%%%%%%%%%%
%% The appropriate \bibliography command should be placed here.
%% Notice that the class file automatically sets \bibliographystyle
%% and also names the section correctly.
%%%%%%%%%%%%%%%%%%%%%%%%%%%%%%%%%%%%%%%%%%%%%%%%%%%%%%%%%%%%%%%%%%%%%
\bibliography{achemso-demo}

\begin{figure}[h!]
\center{\includegraphics[width=1.05\linewidth]{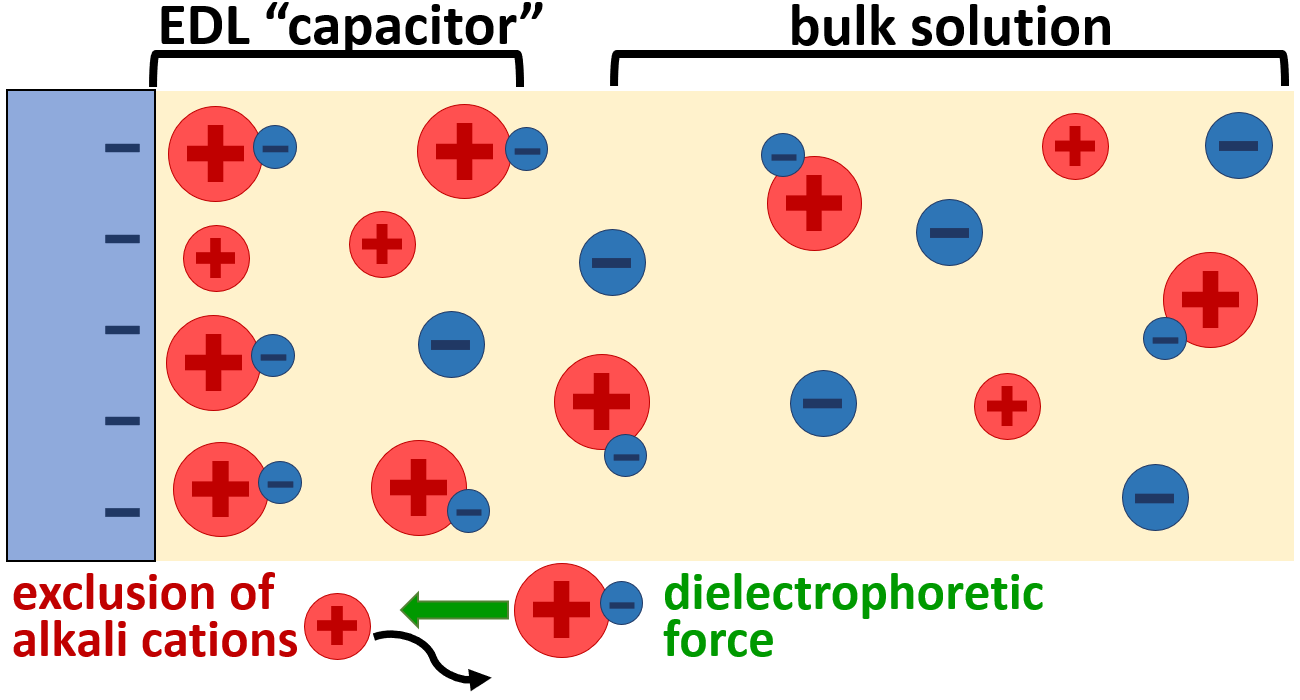}}
\caption{}
\label{Schematic}
\end{figure}

\end{document}